\documentclass[a4paper,11pt]{article}
\usepackage{pos}

\title{Measurement of the jet-particle $v_{2}$ in p--Pb and Pb--Pb collisions at $\sqrt{s_{\rm NN}}$ = 5.02 TeV with ALICE at the LHC}
\ShortTitle{Jet-particle $v_{2}$ in p--Pb and Pb--Pb collisions}

\author*[a,b]{Siyu Tang on behalf of the ALICE Collaboration}

\affiliation[a]{Central China Normal University,\\
  Wuhan, China}

\affiliation[b]{Laboratoire de Physique de Clermont,\\
Clermont-Ferrand, France}

\emailAdd{siyu.tang@cern.ch}

\abstract{The elliptic flow of jet particles at midrapidity ($|\eta|$ < 0.8) is measured for the first time in p--Pb collisions at $\sqrt{s_{\rm NN}}$ = 5.02 TeV and extended down to lower $\it{p}_{\rm T}$ in Pb--Pb collisions at $\sqrt{s_{\rm NN}}$ = 5.02 TeV. The jet particles are extracted with the two-particle correlation method using a two-dimensional fit, and their $v_{2}$ is calculated with the novel three-particle correlation technique. The $v_{2}$ of jet particles is found positive in high-multiplicity p--Pb collisions and exhibits no dependence on $\it{p}_{\rm T}$ within uncertainties. Comparisons with the jet-particle $v_{2}$ in Pb--Pb collisions and the inclusive charged-particle $v_{2}$ in both collision systems are also discussed, aiming to bring new insights into the understanding of the origin of the high-$\it{p}_{\rm T}$ azimuthal anisotropy observed in small collision systems.} 

\FullConference{%
  *** The European Physical Society Conference on High Energy Physics (EPS-HEP2021), ***\\
  *** 26-30 July 2021 ***\\
  *** Online conference, jointly organized by Universität Hamburg and the research center DESY ***
}

\begin{document}
\maketitle

\section{Introduction}
\label{sec:Introduction}
The particles which originate from hard scattering of partons (quarks and gluons), commonly known as hard probes, play an important role in investigating the properties of quark--gluon plasma (QGP) formed in ultrarelativistic heavy-ion collisions. Since the hard scattering processes with a large momentum transfer ($1/Q \ll 1~\rm{fm}/\it{c}$) occur prior to the QGP formation, the hard partons and the subsequent parton shower experience the entire evolution of the system. The angular distribution of final-state particles from hard partons, including high-$\it{p}_{\rm T}$ particles or jets~\cite{Salam:2010nqg}, can be expanded as Fourier series. The second harmonic coefficient, $v_{2}$, refers to the elliptic flow. The elliptic flow is sensitive to the path-length dependent energy loss during the interactions between initial partons and the QGP medium~\cite{ALICE:2015efi}, which is the so-called jet quenching~\cite{Bjorken:1982tu}. However, recent measurements show also a non-zero $v_{2}$ value for high-$\it{p}_{\rm T}$ charged particles at high multiplicities in p--Pb collisions~\cite{ATLAS:2019vcm}, but no jet quenching effect is observed in small systems~\cite{ALICE:2018vuu, ALICE:2017svf}. To understand the origin of such collectivity, the ALICE Collaboration measured the $v_{2}$ of charged particles in jets in p--Pb collisions at $\sqrt{s_{\rm NN}}$ = 5.02 TeV, which allow us to further separate hard and soft components of collectivity in small collision systems. The $v_{2}$ of jet particles in Pb--Pb collisions at $\sqrt{s_{\rm NN}}$ = 5.02 TeV is also measured and extended to lower $\it{p}_{\rm T}$ compared to previous measurements at the LHC~\cite{ALICE:2015efi} in order to provide a baseline for collectivity of hard probes in small systems.

The analysis is carried out with ALICE apparatus~\cite{ALICE:2014sbx} using the main sub-detectors: the Time Projection Chamber (TPC), the Forward Multiplicity Detectors (FMD), the V0, and Silicon Pixel Detector (SPD). The TPC used for charged-particle tracking covers the pseudorapidity range $|\eta|<0.8$ and has a full azimuthal acceptance. The FMD is located at 1.7 $<\eta<$ 5.0 (FMD1,2) and -3.4 $<\eta<$ -1.7 (FMD3) with full azimuthal coverage. The FMD1,2 is employed in p--Pb collisions for the event selection and to construct long-range correlations with TPC tracks to extract the $v_{2}$ coefficient. The V0 detector, made of two scintillator arrays, is used for triggering, event selection, and event activity determination. In addition, the V0 is also used to calculate the flow vector in Pb--Pb collisions, together with the reconstructed tracklets\footnote{The track segments formed by the clusters in the two SPD layers and the primary vertex} in the SPD detector.

\section{Analysis Strategy}
\label{sec:Analysis Strategy}
\subsection{Extraction of the jet signal}
\label{sec:Extraction of jet signal}
\begin{figure}[htbp]
\centering{}
\includegraphics[width=.35\columnwidth]{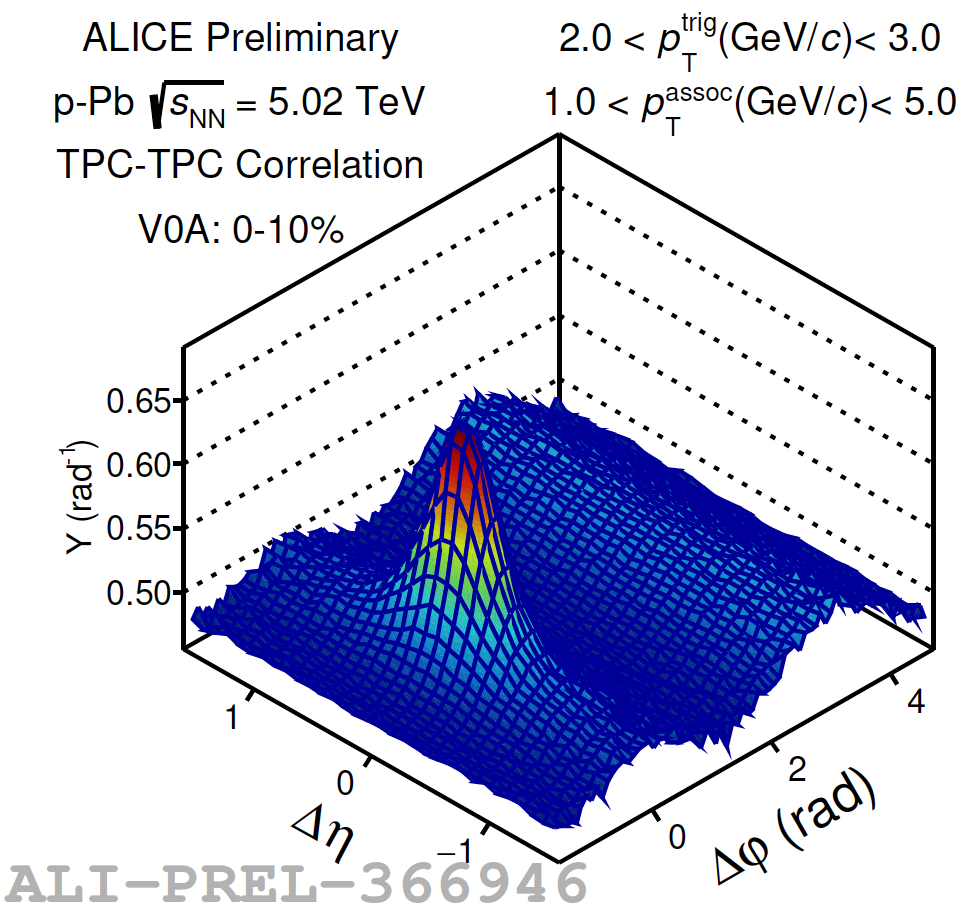}
\includegraphics[width=.35\columnwidth]{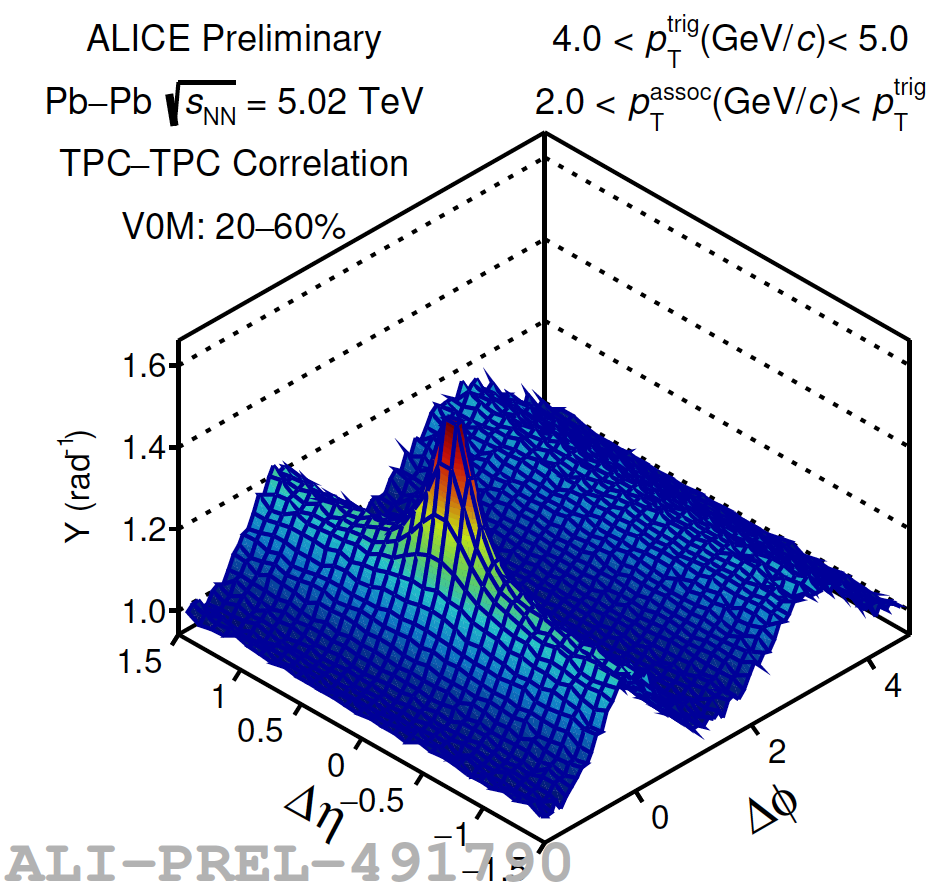}
\caption{The associated yield per trigger particle in TPC-TPC correlation for p--Pb (left) and Pb--Pb (right) collisions at $\sqrt{s_{\rm NN}}$ = 5.02 TeV.}
\label{Jet peak in p-Pb and Pb-Pb}
\end{figure}

\begin{figure}[htbp]
\centering{}
\includegraphics[width=.70\columnwidth]{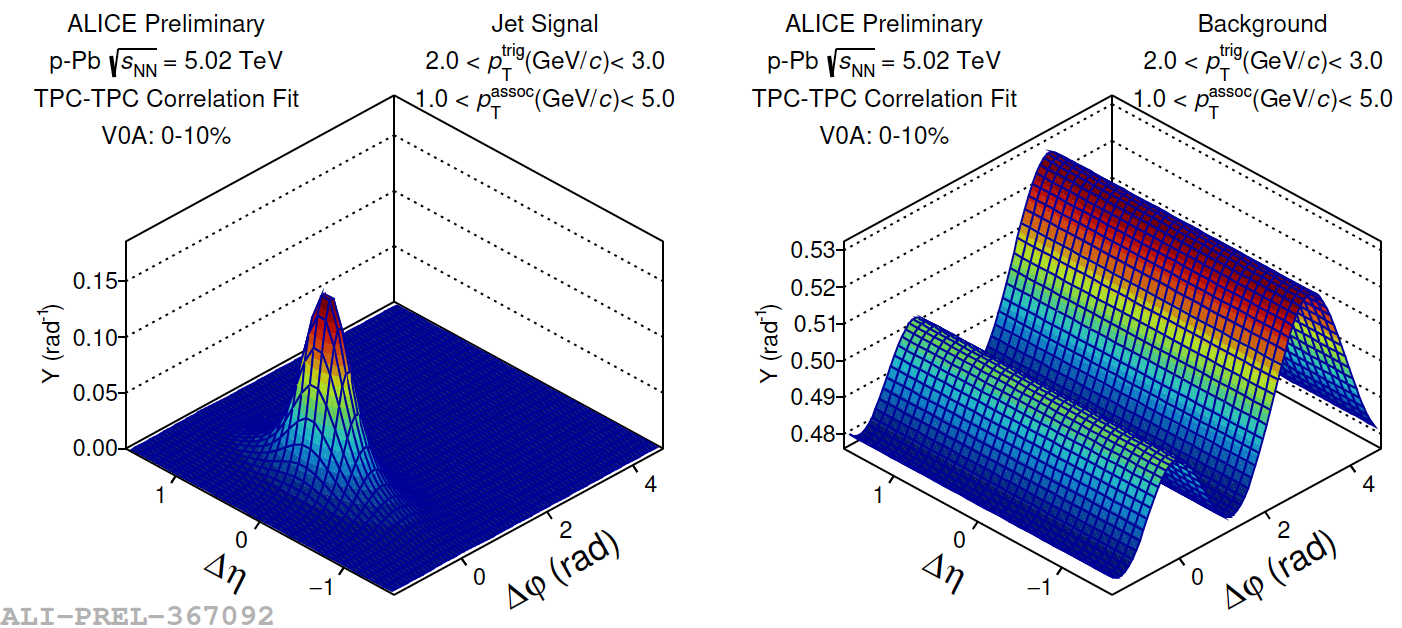}
\caption{The extracted near-side jet (left) and background (right) yields after the fit of TPC-TPC correlation distribution in p--Pb collisions at $\sqrt{s_{\rm NN}}$ = 5.02 TeV}
\label{Jet peak and background in p-Pb}
\end{figure}

A reconstruction algorithm is usually employed to identify jets, but here the jet particles are extracted via two-particle correlations. Both trigger and associated particles are chosen from TPC tracks, and only the same-sign charged particles are selected to suppress the resonances. The associated yield per trigger particle as a function of the pseudorapidity difference $\Delta \eta$ and azimuthal angle difference $\Delta \varphi$ between the trigger and associated particles is defined as
\begin{equation}
  Y(\Delta \varphi,\Delta \eta) = \frac{1}{N_{\mathrm{trig}}}\frac{\mathrm d^{2}N_{\rm assoc}}{\mathrm d\Delta \varphi \mathrm d\Delta \eta} = \frac{S(\Delta \varphi, \Delta \eta)}{B(\Delta \varphi, \Delta \eta)},
  \label{eq:definition of correlation}
\end{equation}
where $\it{N}_{\rm{trig}}$ is number of trigger particles, $S(\Delta \varphi, \Delta \eta) = \frac{1}{\it{N}_{\rm trig}}\frac{\rm d^{2}\it{N}_{\rm same}}{\rm d\Delta\varphi \rm d\Delta \eta}$ is the associated yield per trigger in the same event, and $B(\Delta \varphi, \Delta \eta) = \alpha \frac{\rm d^{2}\it{N}_{\rm mixed}}{\rm d\Delta\varphi \rm d\Delta \eta}$ is the pair yield in different events obtained with the event-mixing technique~\cite{ALICE:2012eyl} and normalized to unity at maximum. By dividing $S(\Delta \varphi, \Delta \eta)$ by $B(\Delta \varphi, \Delta \eta)$, the finite-acceptance effect is accounted for~\cite{ALICE:2015lpx}. The corrected associated yield per trigger particle is shown in the Fig.~\ref{Jet peak in p-Pb and Pb-Pb}, where the jet peak is clearly observed in the near-side region for both p--Pb (left) and Pb--Pb (right) collisions. The yield associated to the near-side jet peak is extracted with a two-dimensional fit method, using a double Gaussian on the near-side superimposed on the sum of harmonics up to fifth order~\cite{ALICE:2014mas}. Figure~\ref{Jet peak and background in p-Pb} shows the extracted jet-peak yield (left) and background yield\footnote{The background corresponds to the particles which are not from the near-side jet peak.}~(right) in p--Pb collisions, which will be used to extract the $v_{2}$ of jet particles in the next step.

\subsection{Calculation of jet-particle $v_{2}$}
\label{sec:Calculation of jet-particle v2}
\begin{figure}[htbp]
\centering{}
\includegraphics[width=.90\columnwidth]{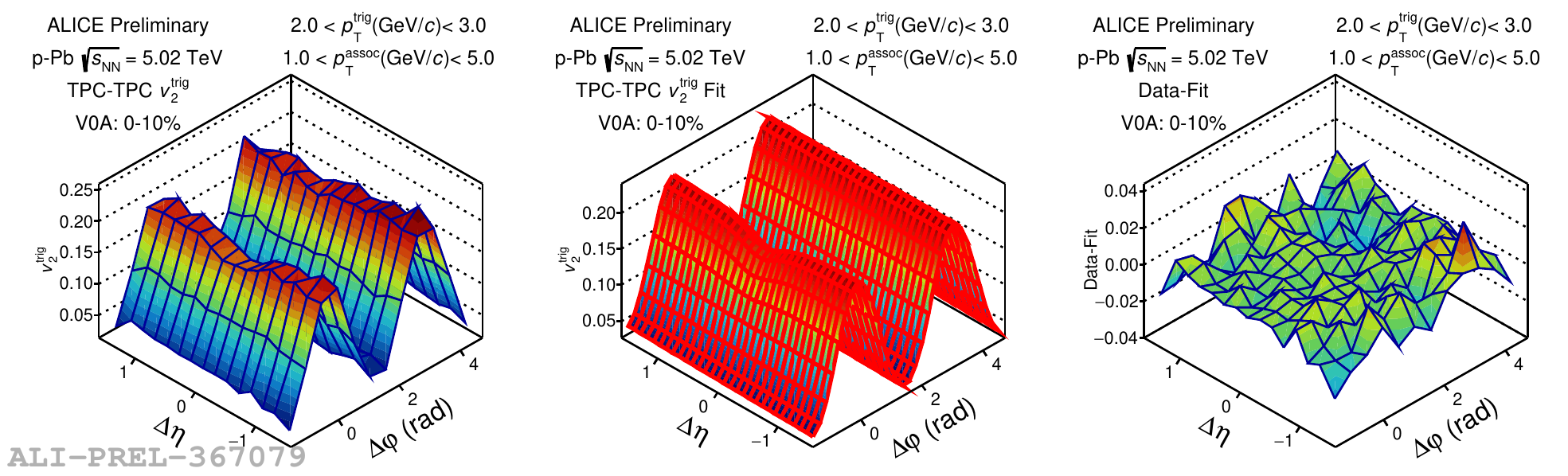}
\caption{Left: $v_{2}$ distribution of trigger particles in TPC-TPC pairs in p--Pb collisions at $\sqrt{s_{\rm NN}}$ = 5.02 TeV. Middle: A fit of Eq.~\ref{eq:fit v2} to $v_{2}$ distribution. Right: difference between data and fit.}
\label{v2 distribution}
\end{figure}

The calculation of the jet-particle $v_{2}$ is based on the so called "three-particle correlation" technique. In the three-particle correlations, two of them are chosen from the particle pair obtained in TPC-TPC correlations, as shown in Fig.~\ref{Jet peak in p-Pb and Pb-Pb}, and the third particle is selected in the forward rapidity region. The long-range correlations between the trigger particle of the TPC-TPC particle pairs and the forward particles are constructed. In p--Pb collisions, the forward particles are chosen from the "FMD1,2" and the flow coefficient is extracted from the Fourier expansion of the long-range correlation after the subtraction of the scaled low-multiplicity class ((0-10\%)-(60-100\%)) to remove the non-flow contribution~\cite{ALICE:2015lpx}. In Pb--Pb collisions, the forward particles are measured in V0A, and the $v_{2}$ is calculated directly by means of the scalar product method with the three sub-event technique~\cite{ALICE:2018rtz}. Figure~\ref{v2 distribution} (left) shows the $v_{2}$ distribution as a function of $\Delta\varphi$ and $\Delta\eta$ of TPC-TPC pairs. The concavity is observed in ($\Delta\varphi\sim0,\Delta\eta\sim0$) where the jet peak is located. This indicates that the jet-particle $v_{2}$ is different from the $v_{2}$ of the background particles. Considering that $v_{2}(\Delta\varphi, \Delta\eta)$ is the weighted sum of the $v_{2}$ from jet particles ($v_{\rm 2,jet}$) and background ($v_{2,\rm bkg}$), where the weight is the ratio of the yield of jet-peak ($\rm Y_{jet}$) to the background ($\rm Y_{bkg}$) obtained when fitting the TPC-TPC correlation, it can be written as
\begin{equation}
v_{2}(\Delta\varphi, \Delta\eta) = \rm Y_{jet}/ (\rm Y_{jet} + \rm Y_{bkg}) \times \it{v}_{\rm 2,jet} + \rm Y_{bkg}/ (\rm Y_{jet} + \rm Y_{bkg}) \times \it{v}_{\rm 2, bkg}(\rm \Delta\varphi, \Delta\eta), 
\label{eq:fit v2}
\end{equation}
where $v_{\rm 2,jet}$ is a constant, and $v_{\rm 2,bkg}$ is a Fourier series up to the fifth order and has a dependence on $\Delta\varphi$ and $\Delta\eta$. The fit of $v_{2}(\Delta\varphi, \Delta\eta)$ distribution is obtained with Eq.~\ref{eq:fit v2} in each $\it{p}_{\rm T}$ interval of trigger and associated particles, and it is shown in the middle panel of Fig.~\ref{v2 distribution}. The small difference between the fit and the data shown in the right panel of Fig.~\ref{v2 distribution} demonstrates that the fit strategy is suited to extract $v_{\rm 2,jet}$.

\section{Results}
\label{sec:Results}
\begin{figure}[htbp]
\centering{}
\includegraphics[width=.45\columnwidth]{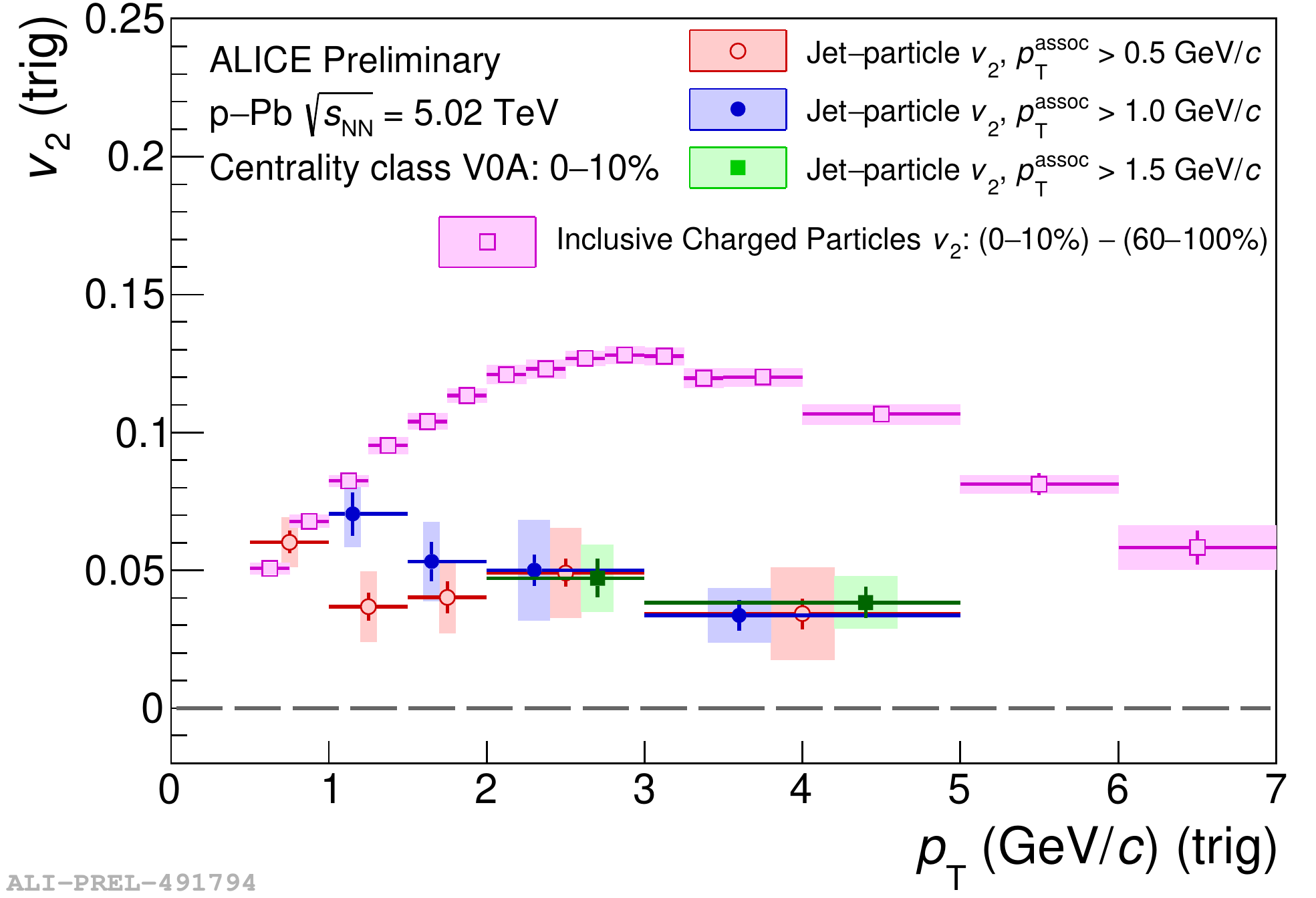}
\includegraphics[width=.45\columnwidth]{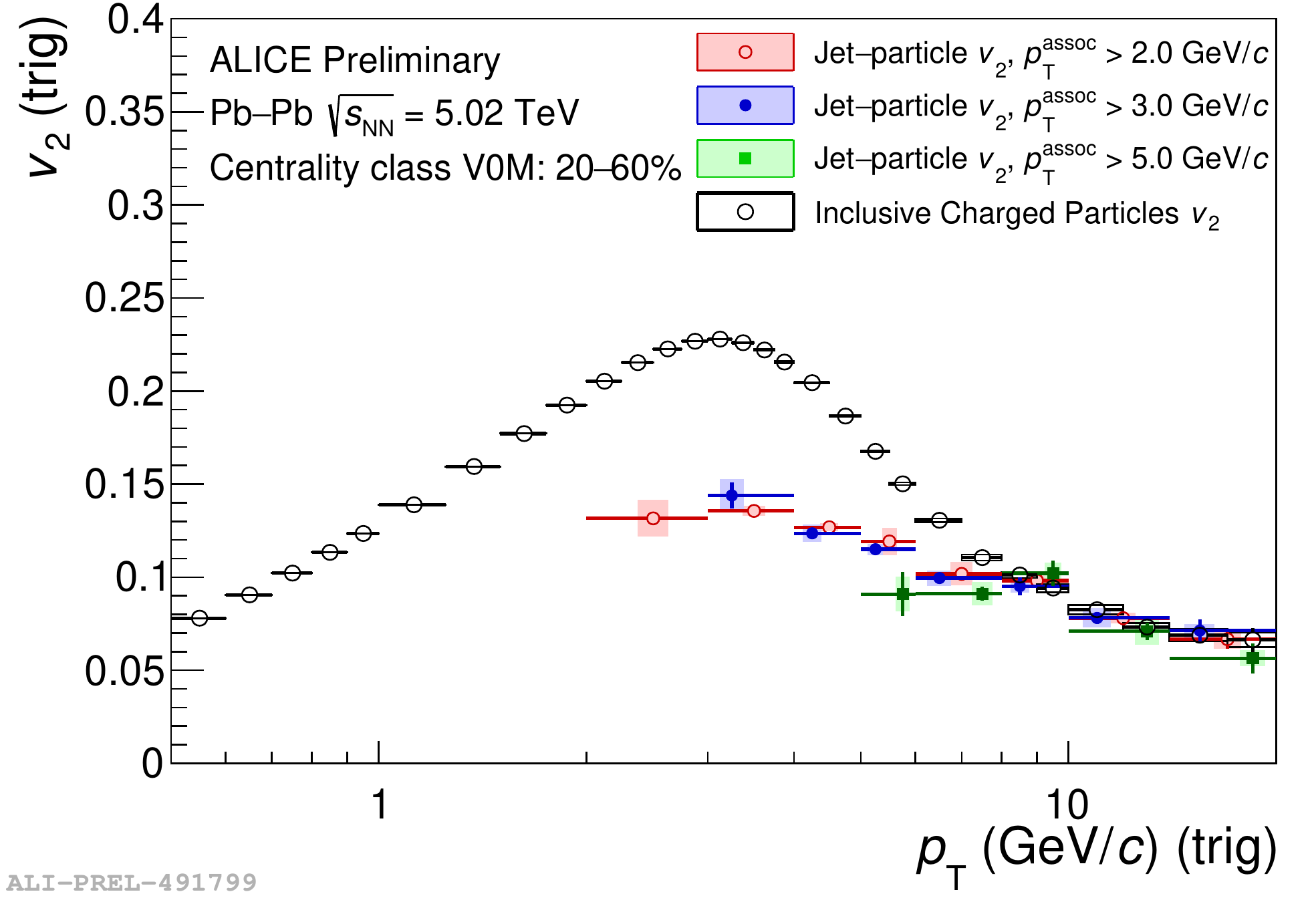}
\caption{The jet-particle $v_{2}$ as a function of the trigger-particle $\it{p}_{\rm T}$ with different associated-particle $\it{p}_{\rm T}$ selections, compared to the inclusive charged-particle $v_{2}$, in p--Pb (left) and Pb--Pb (right) collisions at $\sqrt{s_{\rm NN}}$ = 5.02 TeV.}
\label{Jet v2}
\end{figure}

\begin{figure}[htbp]
\centering{}
\includegraphics[width=.60\columnwidth]{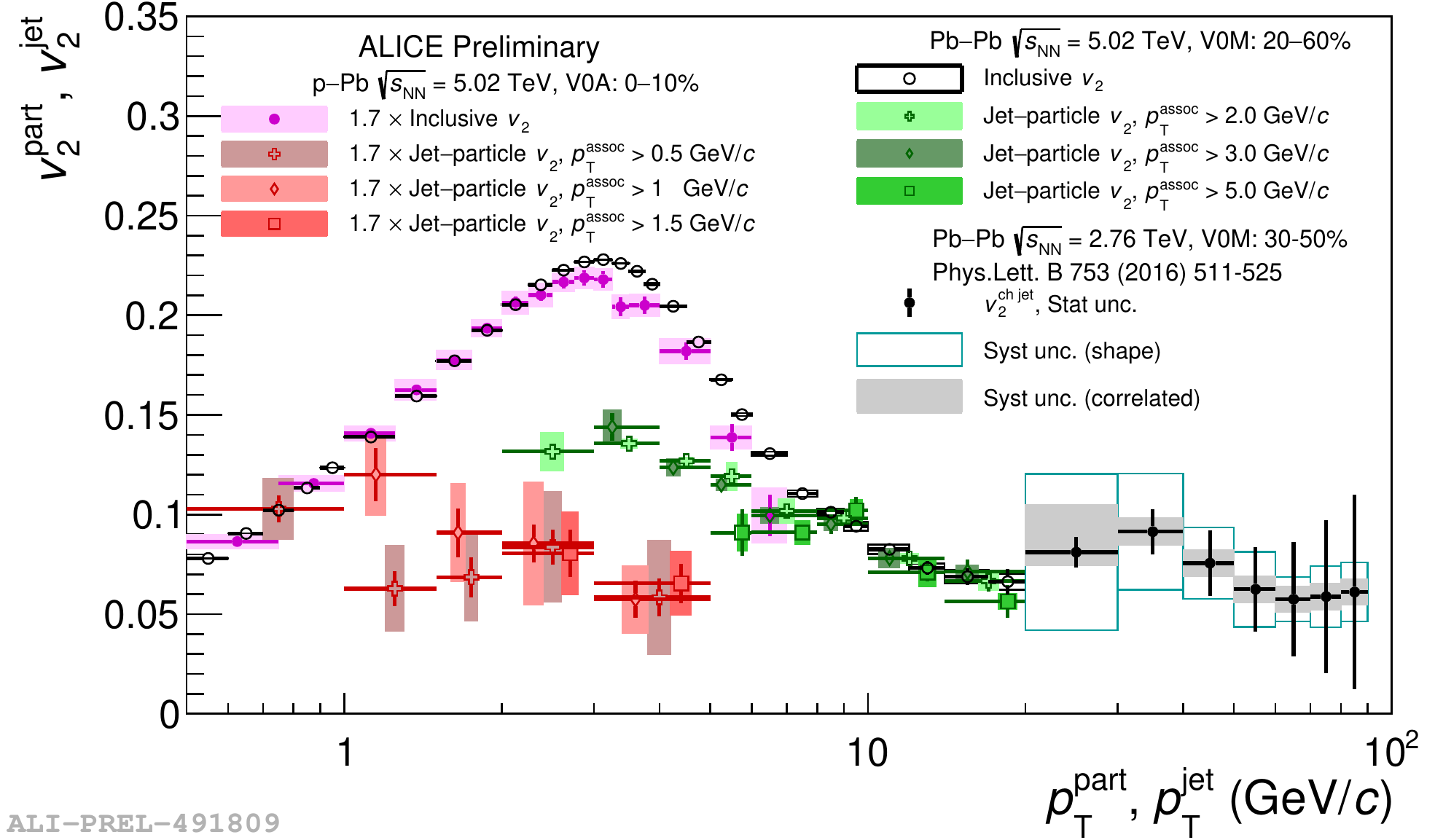}
\caption{Comparison of jet-particle $v_{2}$ and inclusive charged-particle $v_{2}$ in in p--Pb and Pb--Pb collisions at $\sqrt{s_{\rm NN}}$ = 5.02 TeV, and reconstructed-jet $v_{2}$ in Pb--Pb collisions at $\sqrt{s_{\rm NN}}$ = 2.76 TeV}
\label{Jet v2 Comp}
\end{figure}

A positive jet-particle $v_{2}$ is measured for the first time in high-multiplicity p--Pb collisions at $\sqrt{s_{\rm NN}}$ = 5.02 TeV. The magnitude is lower than the $v_{2}$ of inclusive charged particles, as shown in Fig.~\ref{Jet v2} (left). A consistent $v_{2}$ is measured with different trigger- and associated-particle $\it{p}_{\rm T}$ selections. In Pb--Pb collisions, the jet-particle $v_{2}$ is measured for the first time down to 2 GeV/$c$ in 20–60\% Pb–Pb collisions at $\sqrt{s_{\rm NN}}$ = 5.02 TeV, and does not show any dependence on the associated-$\it{p}_{\rm T}$ selections, as shown in Fig.~\ref{Jet v2} (right). Furthermore, the jet-particle $v_{2}$ converges towards the $v_{2}$ of inclusive charged particles in Pb--Pb collisions for $\it{p}_{\rm T}>$~7 GeV/$c$, where the parton energy loss is dominant~\cite{ALICE:2018vuu}. Figure~\ref{Jet v2 Comp} shows the final comparison among the jet-particle $v_{2}$, inclusive charged-particle $v_{2}$, and reconstructed-jet $v_{2}$~\cite{ALICE:2015efi} in p--Pb and Pb--Pb collisions at $\sqrt{s_{\rm NN}}$ = 5.02 TeV. The jet-particle $v_{2}$ in Pb--Pb collisions is consistent with the reconstructed-jet $v_{2}$ at high $\it{p}_{\rm T}$, which are both interpreted by the path-length dependent energy loss effect. On the other hand, in order to compare the results obtained in p--Pb and Pb--Pb collisions directly, the $v_{2}$ of inclusive charged particles in p--Pb is multiplied by a factor 1.7, which is mainly influenced by the initial-state eccentricity~\cite{ATLAS:2019vcm}, to match the $v_{2}$ of inclusive charged particles in Pb--Pb collisions. The same factor is also applied to the jet-particle $v_{2}$ in p--Pb collisions. After the scaling, the $v_{2}$ of jet particles in p--Pb has a magnitude comparable to the jet-particle $v_{2}$ and inclusive charged-particle $v_{2}$ in Pb--Pb collisions at high $\it{p}_{\rm T}$. This hints to a similar collective behaviour of hard probes in large and small collision systems, even there is no jet quenching effect observed in p--Pb collisions.

\section{Summary}
\label{sec:Summary}
The $v_{2}$ of particles in jets is measured, for the first time, in p--Pb collisions and extended to low $\it{p}_{\rm T}$ in Pb--Pb collisions. A non-zero and $\it{p}_{\rm T}$-independent jet-particle $v_{2}$ is found in p--Pb collisions which gives insight into the origin of collectivity in small collision systems. In Pb--Pb collisions, thanks to the low-$\it{p}_{\rm T}$ region reached for the first time for the jet-particle $v_{2}$ measurement, a clear picture of the $\it{p}_{\rm T}$-dependent $v_{2}$ of the jet particles can be established and more differential information to constrain the hard parton in-medium interacting mechanisms is provided. 

\section{Acknowledgement}
This work is supported by the National Natural Science Foundation of China (Grant No. 11775097, 11805079, 12061141008 and 12175085) and national key research and development program of China under Grant (No. 2018YFE0104700).

\bibliographystyle{JHEP}
\bibliography{Mybib}
%\begin{thebibliography}{99}
%\bibitem{...}
%....

%\end{thebibliography}

\end{document}